\documentclass[a4paper]{article}

\usepackage{bbding}
\usepackage{pifont}
\usepackage{amssymb}
\usepackage{graphicx}

\usepackage{longtable}
\usepackage[english]{babel}
\usepackage[utf8x]{inputenc}
\usepackage[T1]{fontenc}

\usepackage[a4paper,top=3cm,bottom=2cm,left=2cm,right=2cm,marginparwidth=1.75cm]{geometry}

\usepackage{array}
\newcolumntype{P}[1]{>{\centering\arraybackslash}p{#1}}
\newcolumntype{M}[1]{>{\centering\arraybackslash}m{#1}}
\usepackage{floatrow}
\usepackage{threeparttablex}

\usepackage{amsmath}

\usepackage[colorlinks=true, allcolors=blue]{hyperref}
\usepackage{adjustbox,longtable,multirow}
\title{A Review on Bio-Cyber Interfaces for Intrabody Molecular Communications Systems}
\author{Yevgeni Koucheryavy, Anastasia Yastrebova, Daniel P. Martins, Sasitharan Balasubramaniam
\thanks{Yevgeni Koucheryavy is with the Tampere
University, Tampere, Finland, 33720, and with the with Higher School of Economics, Moscow, Russian Federation, 101000. E-mail: yevgeni.koucheryavy@tuni.fi}
\thanks{Anastasia Yastrebova is with the VTT Technical Research Centre of Finland, Oulu, Finland, 90570. E-mail: anastasia.yastrebova@vtt.fi}
\thanks{Daniel P. Martins and Sasitharan Balasubramaniam are with the Walton Institute for Information and Communication Systems Science, Waterford Institute of Technology (WIT), Waterford, Ireland, X91 P20H. E-mail: \{daniel.martins, sasi.bala\}@waltoninstitute.ie.}
\thanks{This work was funded by Science Foundation Ireland and the Department of Agriculture, Food, and Marine via the VistaMilk research centre (grant no. 16/RC/3835).}}

\begin{document}

\maketitle

\begin{abstract}
The recent advancements in bio-engineering and wireless communications systems have motivated researchers to propose novel applications for telemedicine, therapeutics and human health monitoring. For instance, through wireless medical telemetry a healthcare worker can remotely measure biological signals and control certain processes in the organism required for the maintenance of the patient's health state. This technology can be further extended to use Bio-Nano devices to promote a real-time monitoring of the human health and storage of the gathered data in the cloud. This brings new challenges and opportunities for the development of biosensing network, which will depend on the extension of the current intrabody devices functionalities. In this paper we will cover the recent progress made on implantable micro-scale devices and introduce the perspective of improve them to foster the development of new theranostics based on data collected at the nanoscale level.
\end{abstract}

\section{Introduction}

The historical advancements of Biomedical Engineering has largely been attributed to solutions developed from multiple disciplines, which has provided new approaches towards monitoring the human health and improving disease diagnostics. For example, communications engineering has been applied to support the development of intrabody devices that are capable to communicate data recordings to external entities. Through these medical wireless telemetry systems cellular processes of interest can be measured and controlled remotely. Despite the current number of studies in this field, these solutions are hindered by their connectivity limitations \cite{kiourti2014implantable, nieto2017evaluating}. This brings opportunities and challenges for the development of novel remote disease diagnostics and therapeutics. One approach to face these challenges has been the design of novel smaller intrabody devices (going to the nanoscale) with enhanced communications capabilities.  Moreover, an improvement on the efficiency of health monitoring it is expected, as these new devices will be able to interconnect among themselves and use the cloud infrastructure to provide full-time access to all the data gathered by them. 

Going into the future, the telecommunications systems (6G and beyond) are expected to be more pervasive and Internet-dependent \cite{zhang20196g}. This will also reflect on wireless medical telemetry systems, which will be composed by several nanoscale sensors that will be deployed into the human body to provide a more sensitive health monitoring and treat any detected disease in real-time. In this scenario, the conventional communications systems will be replaced by a new paradigm based on the exchange of information in the form of molecules, known as {\bf molecular communications}, which will be applied for the sensing at molecular level and the nanoscale intrabody devices interconnections. To communicate with external networks, including the Internet, these systems will require a translator device (i.e., bio-cyber interface) that will convert any molecular signal into electrical, which will continue to be the applied for macro level computer networks, such as the Internet \cite{akyildiz2015internet}. The bio-cyber interface will also convert the different types of detected molecular signals to interface the exchange of information between the different nanonetworks placed inside the human body \cite{akyildiz2015internet}. This novel framework will enable the remote measurement of molecular signals to provide the required network infrastructure for the future design of novel therapeutics.

In this Review, we survey the past and current designs of intrabody nanoscale devices that enable the remote assessment of human's health state and explore their potential to interface with the Internet. We first contextualize the communications aspects required to interconnect intrabody nanoscale devices. Then, we explore the historical evolution of intrabody nanoscale devices and examine their potential to interface molecular level information with conventional computer networks for remote monitoring of human's health state. Lastly, we discuss the challenges that need to be addressed to build a network of intrabody nanoscale devices that interface with the Internet and present the future perspectives for this research field.

\begin{figure}[t!]
\centering
  \includegraphics[width=\textwidth]{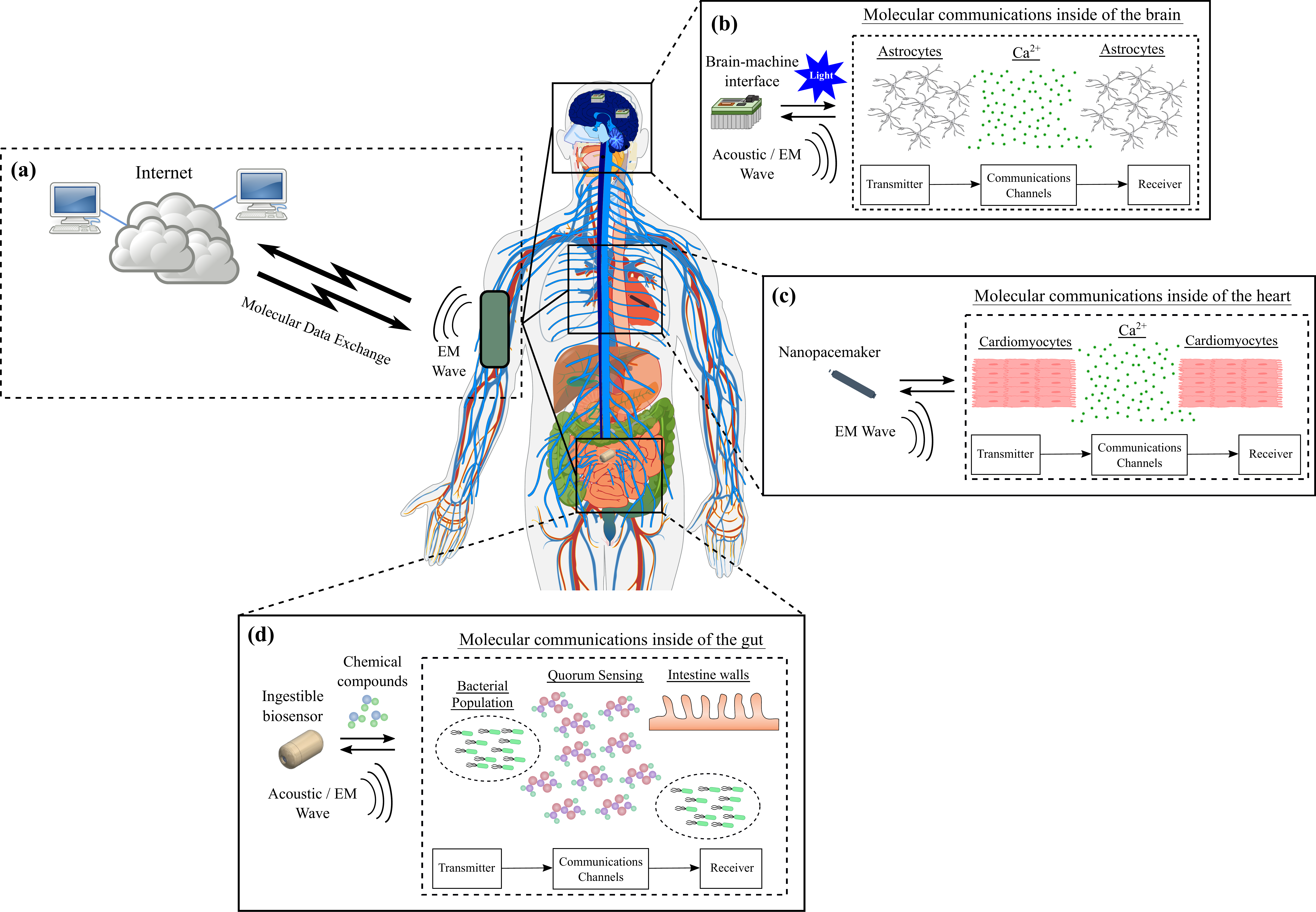}
\caption{Representation of the variety of molecular communications systems that can be placed inside the human body and interconnected using bio-cyber interfaces. (a) The bio-cyber interface exchanges molecular data with the Internet to enable the remote monitoring and control of intrabody devices. (b) brain-machine interfaces use light, acoustic or electromagnetic (EM) waves to stimulate the molecular communications inside of the brain. (c) An implanted nanopacemaker use EM waves to stimulate the heart cells to exchange calcium ions. (d) An ingestible biosensor use chemical signalling, acoustic or EM waves to stimulate a bacteria population to produce quorum sensing signals that can affect the intestine walls and other bacterial populations.}
\label{fig:bio-cyberinterface}
\end{figure}

\section {Internet of Bio-Nano Things}

In the recent years, a novel network paradigm has been developed to integrate nanoscale devices through their communications capabilities to perform complex tasks, such as sensing, actuation and processing of molecular information, namely Internet of Bio-Nano Things. For example, harmful chemical agents dispersed onto the environment can be detected and controlled through this process, as well as, the sensing and actuation on intrabody (of humans, animals and plants) systems through their own health-related information. To enable the Internet of Bio-Nano Things, engineers have been using synthetic biology tools to design cell-based nanoscale devices that are similar to the electronic-based ones deployed on conventional Internet of Things (IoT) scenarios. Through this method, they build nanonetworks that integrate and interconnect those cell-based nanoscale devices through molecular communications systems. 

\subsection{Molecular Communications}

In a molecular communications system, information is produced, propagated and received by cell-based nanoscale devices \cite{el2017engineering}. For example, cells can produce proteins and nucleic acids, i.e. molecular information, and freely diffuse or transport them by exosomes (small membrane vesicles of endocytic origin) for the stimulation of cellular immune responses \(in~vivo\) \cite{thery2002exosomes}. Due to the diversity of natural systems that can inspire the design of these molecular communications systems, different natural cellular signalling processes have been investigated in the past twenty years, such as action potential and quorum sensing, and the characteristics of those natural systems define the requirements for the nanonetwork architecture \cite{dinc2017theoretical}. Therefore, nanoscale devices, that are specific for each system, are built to encode and transmit molecular signal through a communications channel (e.g. fluid media) towards a nanoscale receiver that will retrieve the original molecular information. To encode a molecular information, scientists can change the structure of the molecule, the gene sequence information or the concentration of information in the propagated molecular signal. At the receiver side, the molecular signal is decoded resulting on the extraction of the original information produced by the nanoscale transmitter \cite{nakano2012molecular,wang2017diffusion}. These processes are representations of the chemical reactions required to produce, propagate and receive molecular signals within a nanonetwork. 

Molecular communications systems can be designed to operate using single or a small population of cells depending on the complexity of the desired application. This means that the design parameters of these systems are dependent on the population size of the cells of interest \cite{riglar2018engineering}. For example, each cell of a bacterial population can be engineered to carry the same molecular information between nanoscale devices. At the same time, a whole cellular population can be engineered to behave collectively as a logic gate and process molecular signals \cite{riglar2018engineering}. These both approaches introduce big opportunities for disease treatment, health monitoring, drug delivery, tissue engineering or even DNA engineering due to development of micro-electromechanical systems (MEMS), nanoparticles and other nanoscale technologies. Nevertheless, to be able to remote monitor and control nanoscale processes, an additional device is required, namely bio-cyber interface (see Figure \ref{fig:bio-cyberinterface}). This nanoscale device translates the chemical reactions outputs into other molecular or electric signals that can be seamlessly understood by the conventional computer networks and vice versa (see Figure \ref{fig:bio-cyberinterface}). Therefore, bio-cyber interfaces can be created for any types of cell stimulation or chemical content sensing. In this way, nanonetworks can be deployed inside an animal's body for long-term remote health monitoring, diagnosis and treatment of diseases. By sensing the chemicals of one particular cell and possibility of influence on the cell, it will provide timely prevention and treatment of any disorders such as early inflammation, tumors, and neurological disorders like epilepsy, Parkinson’s and Alzheimer’s disease. 

\subsection{Devices for Internet of Bio-Nano Things}

At the edge of any Internet of Bio-Nano Things network, see Figure \ref{fig:bio-cyberinterface}, there are nanoscale devices that detect and measure molecular signals to inform the bio-cyber interface. These biosensors can detect both physical and chemical stimuli produced by the subject of interest (i.e., environment or cells) and convert it to a signal that is understood by the bio-cyber interface. Healthcare platforms, such as point-of-care diagnostics, have been built using nanoscale biosensors to enable a faster detection of diseases \cite{ghafar2015wireless}. Often, these devices are composed by electrochemical biosensors that would detect chemical signals (e.g., biomolecules) and convert them into electrical signals, that can be stored in the platform's memory and also displayed as the result of the chemical measurement \cite{nagabooshanam2019electrochemical,eissa2017aptamer}. 

Apart from the use of electrochemical sensors to detect biomolecules, biosensors can be designed using enginereed whole cells, at the individual or population level \cite{kylilis2019whole,chen2017development}. These whole cell biosensors provide a simple and cost-efficient solution for the detection and measurement of a wide range of analytes, if compared to the electrochemical sensors, including physical and biochemical signals (e.g., temperature and pH, respectively). Whole cell biosensors are designed using synthetic biology (which is the formal method for the design of artificial systems using biological components) and can be applied to detect and treat cancers, assess the health risks associated to environmental pollution, and discover novel antibiotics, for example. 

\section{Evolution of Intrabody Nanoscale Devices} 
First bio-compatible sensor systems were proposed by Leland C. Clark and C. Lyons for measuring the pH, $p\text{CO}_2$ and $p\text{O}_2$ in the patient's blood, in 1962 \cite{clark1962electrode}. Five years later, Wyrwicka and Stermanin introduced an electronic system to control adult cats' brain signals  \cite{wyrwicka1968instrumental, chaudhary2016brain}. Since then, the sensor design has been improving in term of size, weight, area of action/sensitivity and lifetime \cite{vigneshvar2016recent}. For instance, the current sensor designs have been applying novel bio-compatible soft materials such as silicon, gold, platinum, iridium, as well as, new nanomaterials (carbon nanotubes or graphene) to obtain a higher resolution of electrophysical signals, when compared with the ones used in the past \cite{lee2016soft, wise2004wireless,chen2013flexible}. Nowadays, the applications for nano-implants vary from the detection and monitoring to the treatment of potential harmful health conditions in the cellular level. Table \ref{tab:table1} shows a list of intrabody nanoscale devices, developed in the past years, with their function and technical characteristics descriptions.

Once implanted in the animal body, external signals are applied to monitor and control the operation of nanoscale devices such as acoustic waves and electromagnetic signals \cite{gil2021low,chow2010fully,majerus2011low,varel2014wireless,rajavi2017rf}. These external signals have  sufficient penetration capability through the animal's skin tissue without affecting it and the other cells surrounding the implanted device. For example, with the help of acoustic waves it is possible to deliver nanoparticles and nanoscale devices to a desired location and keep them there releasing the treatment chemicals, such as drugs, insulin and interferons \cite{lavan2003small}. Furthermore, acoustic and electromagnetic waves can also be applied by those devices to affect the cellular activity that they are monitoring. Researchers have been also able to manipulate the cellular activity through optogenetic stimulation of a cell's light-gated ion channels \cite{amar2015power}. The typical wavelength that is used for stimulation is 460 nm (blue light), as it induces the expression of proteins (channelrhodopsin-2 - ChR2) to gate the neuron membrane and produce electrical spikes.  Other wavelengths, like red (635-735 nm) and yellow (560-590 nm) are also applied to induce other membrane gating proteins \cite{pastrana2010optogenetics}. Therefore, neural activity can be excited or inhibited depending on the type of opsins (light activated proteins) expressed in the target neurons \cite{rossi2015wirelessly}. By varying the temperature around the region of interest, perturbations on the cellular activity originated either from the intrinsic properties of the plasma membrane or from the temperature sensitivity of the opsins can be induced \cite{rossi2015wirelessly}. For instance, spatially localized temperature gradients on neuronal tissues were reported to be responsible for the increase in membrane capacitance and the consequent trigger of action potential firing \cite{colombo2016nanoparticles}. Electric fields can be emitted by the nanoscale devices to stimulate cellular activity, which traditionally is carried out using metal electrodes placed in contact with cells to achieve a capacitive coupling. Another option is use transcranial magnetic cellular stimulation, which couples high intensity magnetic fields to local electrical currents in neural systems \cite{hallett2000transcranial}.

\renewcommand{\arraystretch}{1.6}
\begin{ThreePartTable}
\begin{longtable}{M{0.05\textwidth}M{0.03\textwidth}M{0.1\textwidth}M{0.2\textwidth}M{0.1\textwidth}M{0.08\textwidth}M{0.13\textwidth}M{0.1\textwidth}}
\caption{Bio-compatible intrabody devices timeline, with their respective technical characteristics descriptions.}
\label{tab:table1}\\

\hline
{\bf Ref.} & {\bf Year} & {\bf Size} & {\bf Function} & {\bf Power Supply}  & {\bf Interface} & {\bf External Comms.} & {\bf Mol. signal} \\ \hline
\endfirsthead
\caption{Continued from the previous page.}\\
\hline
{\bf Ref.} & {\bf Year} & {\bf Size} & {\bf Function} & {\bf Power Supply}  & {\bf Interface} & {\bf External Comms.} & {\bf Mol. signal} \\ \hline
\endhead
\cite{clark1962electrode} & 1962 & 90-122 cm\tnote{a} & Intravascular electrodes for continuous blood monitoring & Mines supply & Wired & Representation of physiological information on the wall-mounted panel & pH, $p\text{CO}_2$ and $p\text{O}_2$ \\

\cite{burton1971rf}\tnote{b} & 1971 & $1.6\times10$ mm & Metallic implant for telethermocoagulation of the brain & RF generator operating at 610 kHz & -- & -- & -- \\

\cite{goffinet1980125iodine} & 1980 & 1-3 cm (diam.) & Iodine prostate implant & -- & Wireless & Radiographs for determining the place of implant & --\\

\cite{shichiri1982wearable} & 1982 & 2 cm, 0.4  mm (diam.) & Needle-type glucose monitoring sensor\tnote{c} & -- & Wired & -- & Glucose \\

\cite{clark1982implanted} & 1982 & 8 mm & Glucose sensor & -- & Wired & Information displayed on polarograms & Glucose\\

\cite{shichiri1986telemetry} & 1986 & TX size: $4 \times 6 \times 2$ cm; RX size: $10 \times 12 \times 15\,\text{cm}$ & Glucose monitoring & Lithium battery\tnote{d} & Wireless & Information displayed in LED panel & Glucose\\
   
\cite{turner1990biocompatible} & 1990 & 7 mm (diam.), 0.3 mm\tnote{e} & Continuous glucose monitoring (\emph{in vivo}) & -- & Wired & Output voltage according to measurements & Glucose\\

\cite{moscone1992subcutaneous} & 1992 & 4 mm (diam.),  10 mm\tnote{f} & Subcutaneous probe coupled with a glucose bio-sensor & -- & Wired & Electrochemical detector & Glucose\\

\cite{freaney1997novel} & 1997 & 380 $\mu$m (diam.) & Glucose and lactate real time monitoring & -- & Wired & Computer software & Glucose and lactate \\

\cite{lowry1998amperometric} & 1998 & 5 cm & Extracellular glucose monitoring in brain & -- & Wired & Computer & Glucose\\

\cite{crespi2002vivo} & 2002 & 30 $\mu$m (diam.) & Ischaemia, depression and drug dependence monitoring & -- & Wired & -- & Ascorbic acid, catechol, 5-OH-indole and peptidergic oxidation signals \\

\cite{maloney2005electrothermally} & 2005 & 20 mm, 120 nL & Electro-thermally activated microchip for implantable drug delivery and bio-sensing\tnote{g} & -- & Wired & -- & $^{14}$C-labelled mannitol \\

\cite{justin2009biomimetic} & 2009 & $4 \times 2 \times 0.5$ mm & Biosensing & -- & -- & Computer & Intramuscular lactate, glucose and pH\\

\cite{chow2010fully} & 2010 & $3 \times 6$ mm & Cardiac pressure sensing system\tnote{h} & RF powering & Wireless & Application-specific integrated circuit (ASIC), MEMS sensor for the pressure measurements & -- \\

\cite{majerus2011low} & 2011 & $7 \times 15 \times 4$ mm & Bladder-pressure-sensing system for chronic applications & Quallion micro-battery\tnote{i} & Wireless & External RF receiver–recharger & -- \\

\cite{varel2014wireless} & 2014 & 13 mm (diam.) & Intraocular pressure monitoring\tnote{j} & -- & Wireless & PC & -- \\

\cite{lee2016implantable} & 2016 & $1000 \times 100\mu$m & micro-coil for intracortical magnetic stimulation & External generator & Wired & -- & --\\

\cite{rajavi2017rf} & 2017 & $2 \times 1.6 \times 0.6\,\text{mm}$ & Radio system for neural implants\tnote{k} & RF powering & Wireless & External reader & -- \\\hline
\end{longtable}
\begin{tablenotes}
    \vspace{-0.3cm}
    \scriptsize
    \item[a] Microcatheter size.
    \item[b] First brain implant used clinically
    \item[c] Used with artifical endocrine pancreas (box of 12 x 15 x 6 cm)
    \item[d] Lifetime of 3 days
    \item[e] Thickness
    \item[f] Length
    \item[g] Electro-thermal mechanism to release drugs or expose bio-senors
    \item[h] Data transmission distance of 10--100 cm
    \item[i] Recharge period of 6 h
    \item[j]  The device is powered from a distanmce of 1--2 cm
    \item[k]  Working prototype
    \end{tablenotes}
\end{ThreePartTable}

\section{Using Intrabody Devices as Bio-Cyber Interfaces}\label{sec:intrabodybio}
For the past sixty years, intrabody devices are been applied to control or treat diseases in humans. As their natural evolution, these devices will be interconnected among themselves and even connected to the Internet. Beyond that, they can become bio-cyber interfaces to enable the exchange of any molecular signal between different intrabody Bio-Nano Things networks. Here we focus on specific parts of the human body to examine the potential of this novel role for intrabody devices.

\subsection {Nervous and Neural Interfaces}
In order to monitor and control the brain activity, control units that are able to translate the natural signals from brain into understandable machine commands (i.e., Brain-Machine Interfaces - BMIs) has been developed in the past years \cite{lebedev2006brain}. BMIs are conceived as a new treatment  method or therapy for neural diseases, and cognitive abilities problems \cite{urdaneta2017central}. For instance, patients with the loss of somatosensation, (i.e., tactile sensation) suffer from the serious deficits in motor control or struggle to manipulate objects using prosthesis, can be treated with neurostimulation using BMIs. Recently, significant progress have been made in this area, as showed by the success in restoring tactile sensation and motor control of animals \cite{d2017somatotopic, o2011active, shanechi2014cortical, ganzer2018closed} and humans \cite{collinger2013high, flesher2016intracortical}. Moreover, a neuromotor prosthesis was operated using a BMI implanted in the motor cortex of a 25-years old human who is unable to move or sense his limbs due to tetraplegia \cite{hochberg2006neuronal}. In this case, the patient was able to manipulate external devices of any kind by controlling the computer cursor using its brain. There are also studies and experiments regarding visual cortex implants, where neural stimulation has been applied to develop an artificial vision system \cite{normann2007technology}. 

Despite the success of the therapies using neural interfaces, these old BMIs were bulky and uncomfortable for the everyday usage. Therefore, newer research have been tackling this bottleneck by developing tiny and more comfortable BMI prototypes that are able to stimulate and record the action potentials (i.e., brain electric signals) at single cellular level \cite{rajavi2017rf, hsiao2016organic, kwiat2012highly, hai2010long}. For example, an implantable device, using CMOS technology, was developed for \emph{in vivo} photostimulation, and optical Ca2+ imaging of neurons in the brain \cite{kobayashi2016optical}. The main applications of the device are regenerative medical transplantation and gene delivery to the brain using a photosensitive channel and a fluorescent physiological indicator. The device consists of eight green light emitting diodes (LEDs) for fluorescence excitation, three separate blue LEDs for localized photostimulation, and a CMOS sensor chip for $\text{Ca}^{2+}$ imaging. The sensor chip size is $1\,\text{mm}\,\times\,3.5\,\text{mm}$, whereas the imaging area of an active pixel sensor is $900\,\mu\text{m}\,\times\,2,010\,\mu\text{m}$. The device is connected with external devices (i.e., computer) and can send and read data from the brain (bidirectional link) \cite{kobayashi2016optical}. It is possible to use a wired or wireless transmission to interconnect the BMI with the computer. This BMI was implanted in visual cortex of brain as well as have been tested on cultured Neuro2a cells, obtained from mouse neuroblastoma cells \cite{kobayashi2016optical}. The results indicate that device is capable to perform one specific-targeted cell stimulation, and also showed that an external device could communicate with the cell by its own request, using light, to visualize the results simultaneously to the data transmission.

Other BMIs have been recently proposed to monitor and control the neuron's stimulation. For example, a multielectrode non-Faradaic array (256 × 256 electrodes) CMOS-based with the external trigger was designed to allow a controlled neurostimulation \cite{lei2011high}. This device is capable of stimulate single cells and have a precise control over the stimulation thresholds, which increase the quality of the medical treatment. This CMOS-based microelectrode array is a square electronic sheet of $4\,\text{mm}\,\times\,4\,\text{mm}$, where each individual electrode has an edge length of $11.4\,\mu\text{m}$ and a pitch of $12.2\,\mu\text{m}$ , giving rise to a total active stimulation area of  $3\,\times\,3\,\text{mm}^2$ with an electrode density of $6724\,\text{mm}^{-2}$. The chip contains the high-affinity calcium indicator Fluo-4-AM which perform fluorescence imaging through calcium signaling. Calcium concentration change is presented as an indirect measure of action potential, and the fluorescence signals are recorded simultaneously at 10 Hz and 25 Hz. 

Fluorescence signals is widely applied as indirect measurements of chemical concentrations, but it can also be applied to stimulate neurons. An upconversion-based strategy using nanoparticles was proposed as a combinatorial neural stimulation \cite{lin2017multiplexed}. These spectrum-selective upconversion nanoparticles (UCNPs) absorb near-infrared energy and convert it into visible light for optostimulation of the neuron (ranging from single to group of cells), with the help of channelrhodopsin proteins \cite{lin2017multiplexed}. The device was implanted into the visual cortex of a rat and has the size of $1.5\,\text{mm}\,\times3-5\,\text{mm}$. The UCNPs are covered with glass micro-optrode to become biocompatible, and are excited using a NIR laser (980 nm), which result in green or blue light, depending on the used additive. 

Brain machine interfaces has also been applied to medical diagnostics and therapeutics. For instance, a BMI system was develop to detect an upcoming epileptic seizures and apply an arbitrary-waveform current-mode biphasic stimulation to prevent it. In this case, a neurostimulator was designed to amplify and digitize the detected action potential in a single $\Delta\Sigma$-based neural Analog-to-Digital Converter (ADC) \cite{kassiri201727}. The device has the size of $2.6\,\times\,2.3\,\text{mm}$ and includes 64 closed-loop neurostimulators, a low-power Digital Signal Processor (DSP) with a compact mixed-signal Finite Impulse Response (FIR) filter, two ultra wideband (UWB) transmitters (distances up to $2\,\text{m}$), an inductive power, and a command receiver. The DSP is used to calculate the phase synchrony of the upcoming epileptic seizure, which will trigger the biphasic neurostimulation to prevent it. 

A remote controlled gene-based therapy was proposed using brain machine interface \cite{folcher2014mind}. In this case, the transgene expression of a specific human glycoprotein (secreted alkaline phosphatase - SEAP) in human cells can be triggered by the light emitted from a optogenetic device linked to a brain machine interface that detect the brain activities and mental states \cite{folcher2014mind}. The signaling pathway required for the activation of the SEAP production is based on the NIR light detection system of Rhodobacter sphaeroides bacteria, which produces cyclic diguanylate (c-di-GMP) after being induced by light-sensor proteins (BphG1) \cite{folcher2014mind}. For this application the c-di-GMP is used as the SEAP proteins inducer, and the brain machine interface is designed in form of an electroencephalography (EEG) system that is meant to process state-specific brain waves programs. For the confirmation of remote controlling of transgene expression, hollow-fiber microcontainers containing transgenic cells were implanted into mice, which were transdermally illuminated with a near-infrared light. An EEG-brain set was used to capture the brain waves and identify the specific electric pattern to test this system.

The brain machine interface can be wirelessly interconnected with external devices. For instance, a chip-less wireless neural probe system was proposed for both performing stimulation and reading of neural activities simultaneously \cite{kim2016development}. This device consist of two antennas, which allows to transfer data and wirelessly power the system, and have the following dimensions: $7\,\text{mm}$, $50\,\mu\text{m}$, and $31\,\mu\text{m}$ (length, width, and thickness, respectively). This device have a maximum peak stimulation voltage that can reach to $100\,\mu\text{m}$. Moreover, the device does not sense any chemicals, but detect neuron's electrical pulses. A similar approach was considered to design a microsystem based on electrocorticography and placed it on the surface of the cerebral cortex \cite{muller2015minimally}. This microsystem consists of a 64-channel electrode array and a flexible antenna, and has size of $2.4\,\text{mm}\,\times\,2.4\,\text{mm}$. The device does not influence on the cell, but performs data acquisition, wireless power and data transmission. This device is suitable for long term monitoring due to the use of biocompatible materials.


\begin{table}[t!]
\caption{\label{tab:implants}Characteristics comparison for the current implantable devices.}
\renewcommand{\arraystretch}{1.6}
\begin{tabular}  {M{0.05\textwidth} M{0.20\textwidth} M{0.18\textwidth} M{0.15\textwidth} M{0.18\textwidth} M{0.09\textwidth}} 
\hline
{\bf Device} & {\bf Size} & {\bf Type of influence on the cell} & {\bf Target cell} & {\bf Power / Operating frequency} & {\bf Two-way comms.} \\\hline
\cite{kobayashi2016optical} & 2 × 3.5 mm & Photostimulation & N2a cells & 610 – 670 THz (blue); 540 – 580 THz (green) & Yes\\
\cite{kassiri201727} & 2.6 × 2.3 mm  & Electrostimulation & Brain (mouse model) & -- & -- \\
\cite{lei2011high} & 4 mm × 4 mm & Electrostimulation & Hippocampal neuronal cultures & -- & Yes \\
\cite{kim2016development} & $7\,\text{mm} × 50\,\mu \text{m} × 31\,\mu\text{m}$ & Electrostimulation & Physiological saline solution & 450 MHz & Yes\\
\cite{lin2017multiplexed} & 1.5 mm × 3-5 mm & Light stimulation (NIR) & ChR2 neurons; Cultured hippocampal cells & 1.5 Hz and 10 Hz & No \\
\cite{folcher2014mind} & 23 mm (diam.) & Optogenetic & Stem cells & 430 THz & Yes\\
\cite{muller2015minimally} & chip area 2.4 × 2.4 mm total size 6.5 mm & - & Cerebral cortex & 300 MHz & No\\
\cite{seo2016wireless} & 0.8 × 33 × 31 mm & Electrostimulation & Sciatic
nerve gastrocnemius muscle & 1.85 MHz & Yes\\
\cite{jadhav2012cyborg} &  $\text{diam.} < 100\,\mu \text{m}$ & Photostimulation & Eye & -- & No \\
\cite{kim2013injectable} & $6.45\,\mu \text{m},\, 50 × 50\,\mu \text{m}$ & Optostimulation & HEK293 VTA-DA neurons & 910 MHz & Yes\\
\cite{ho2014wireless} &  2 mm (diam.), 3.5 mm (len.) & Electrostimulation & Heart & 1.6 GHz & No\\
\cite{ye2007remote} & $200\,\mu \text{m}$& Chemical & L929 fibroblast cells & 2-3 W & No\\\hline
\end{tabular}
\end{table}

\subsection {Organs and Soft Tissues Interfaces}

\subsubsection{Ocular Interfaces}

Scientists from the University of California at Berkeley together with the Polytechnic University of Turin, Italy have developed a millimeter-scale probe that is implanted directly into a beetle eye to perform neural stimulation and recording \cite{jadhav2012cyborg}. The size of implanted part this device is less than $100\,\mu \text{m}$ in diameter. To have better performance and to decrease the losses, the device was implanted in the early stage of eye developing of the beetle, because at this stage its neurons can be regenerated, if damaged. The neuron stimulation was done by using blue light emitting diodes, which were placed directly in front of the eye at different distances \cite{jadhav2012cyborg}. The device has showed stable performance over 8 hours of continuous recording and could be repeated for the lifetime of the insect.

\subsubsection {Renal Interfaces}

In the neural engineering field, physiological dysfunctions are approached by identifying the target nerves and providing artificial stimulation to restore the function. For bladder control problems, electrical stimulation has been used as one of the treatments, while only a few emerging neurotechnologies have been used to tackle these problems \cite{lee2015emerging}. For example, a neuroprosthesis (with the size of $100\,\mu \text{m}\,\times\,100\,\mu \text{m}\,\times\,3\,\text{mm}$) was proposed for bladder control \cite{chew2013microchannel}. This neuroprosthetic interface was implanted in spinal chord and attached to the peripheral nerves of a rat model to measure bladder fullness and prevent the spontaneous voiding function by applying a high-frequency conduction block of 20 kHz on the ventral roots \cite{chew2013microchannel}. When needed, the system allows to empty the rat bladder by applying a low-frequency stimulation of  30 Hz. When the bladder became full enough to initiate spontaneous voiding, high frequency activity was detected and the telemeter data about bladder status were sent to the external device.  

\subsubsection{Cardiac Interfaces}

intrabody nanoscale devices have also been designed to interface the measurement of biochemical signals in the heart and circulatory system with external devices. For example, a cardiac interface was proposed to safely wirelessly power an implanted cadiac pacemaker (millimeter scale device) \cite{ho2014wireless}. The device is capable of closed-chest (rabbit) wireless control of the heart as well as it has been tested in porcine brain. The method allows to power nanoscale devices implanted with up to 5 cm of depth. The powering device is placed outside the body, it has dimensions of $6\,\times\,6\,\text{cm}$, and power the cardiac pacemaker using an eletromagnetic signal with the frequency of 1.6 GHz \cite{ho2014wireless}. The device consists of a multiturn coil structure, rectifying circuits for AC/DC power conversion, a silicon-on-insulator integrated circuit (IC) for pulse control, and electrodes to stimulate the heart \cite{ho2014wireless}. Experiments were conducted with a cardiac pacemaker (2 mm diameter and 3.5 mm of height) implanted into the lower epicardium of a rabbit, and its heart rate was monitored through an ECG. The size of the implant 2 mm in diameter, 70 mg and is capable of generating $2.4\,\mu\text{J}$ pulses at rates dependent on the extracted power. The device does not contain a battery, it is powered remotely \cite{ho2014wireless}. A portable, handheld power source was placed 4.5 cm distant from the device, after closing the chest, and it delivered 1 W of power to the cardiac pacemaker. The rabbit's cardiac rhythm was controlled wirelessly by adjusting the operating frequency \cite{ho2014wireless}. This powering system can be applied for any other optical or electrical stimulation task in the body, including neurons or muscle cells.

The circulatory system that surround cancer cells can be monitored using a milimeter scale device that measure the fluid pressure that flows through the area \cite{song2016implantable}. The proposed Interstitial Fluid Pressure Sensor (IFP) enables the daily monitor cancer tumors and send the measurements wirelessly to an external network analyzer \cite{song2016implantable}. The millimiter scale implanted device contains a Guyton chamber which enables an accurate measurement of IFP without interferences from other tissue components. The sensor consists of a coil, an air chamber, a silicone membrane embedded with a nickel plate, and a Guyton chamber. Implant has 3 mm in diameter and 1 mm in thickness and the sensor shows a linear response to the pressure with a sensitivity of 60 kHz/mmHg and a resolution of 1 mmHg (or 133.32 Pa). Continuous tumor IFP monitoring could potentially ensure a timely administration of chemo/radiotherapy agents (as it detects the tumor growth).

\begin{figure}[t!]
\centering
\includegraphics[width=\textwidth]{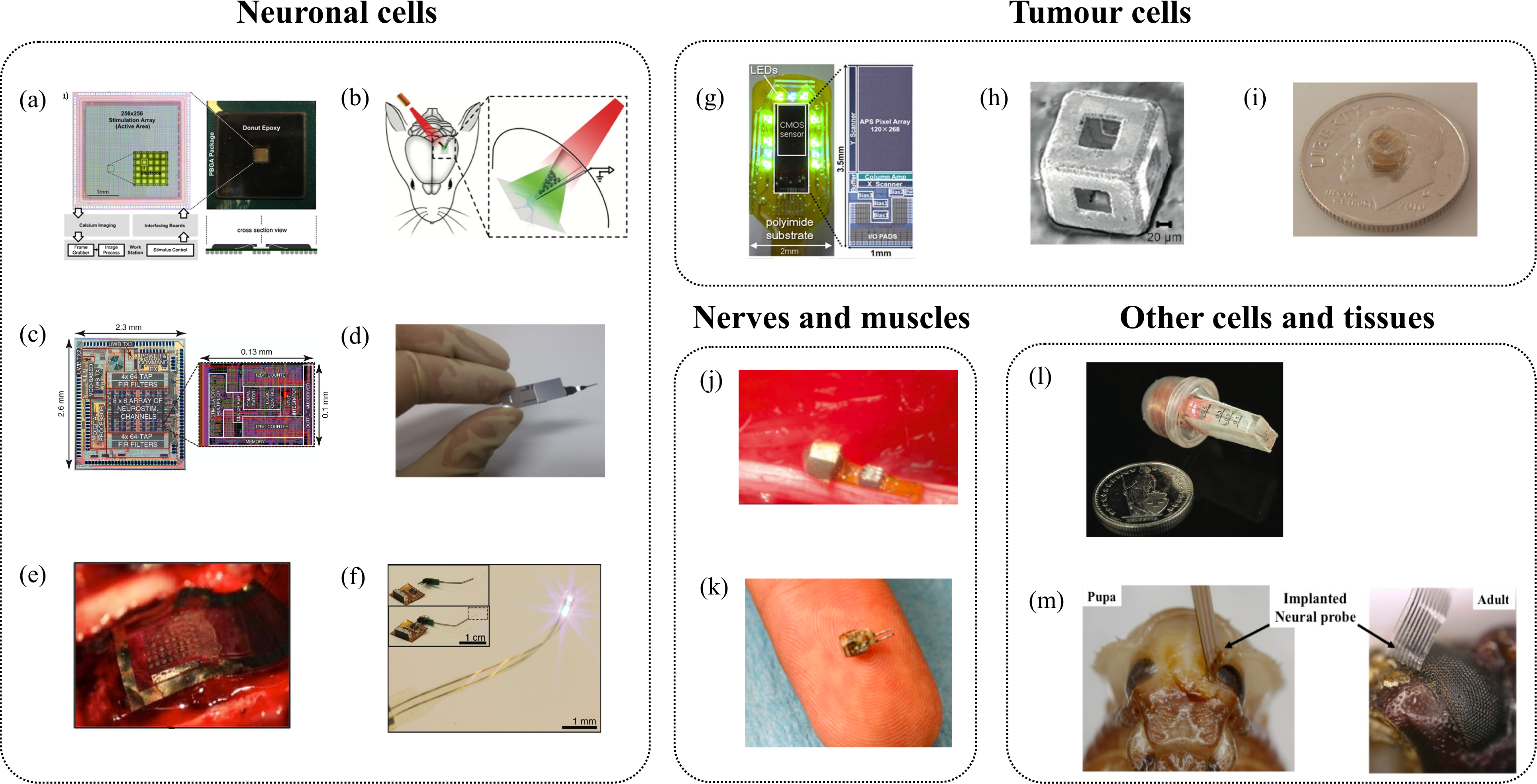}
\caption{Real examples of intrabody devices: (a) hippocampal cell electro-stimulator \cite{lei2011high}; (b) neuronal stimulator and brain signal recorder \cite{lin2017multiplexed}; (c) Wireless neurostimulator \cite{kassiri201727}; (d) neuronal activities recording probe and stimulator \cite{kim2016development}; (e) neuronal signal recorder for cerebral cortex \cite{muller2015minimally}; (f) electro-stimulator for neurons and soft tissues \cite{kim2013injectable}; (g) neuroblastoma cell stimulator \cite{kobayashi2016optical}; (h) drug delivery system \cite{ye2007remote}; (i) tumor monitoring system for prostate \cite{song2016implantable}; (j) monitoring system for muscle and nerve applications \cite{seo2016wireless}; (k) cardiac pacing implant \cite{ho2014wireless}; (l) system for gene expression in human cells \cite{folcher2014mind}; (m) Eye probe for neuronal stimulation \cite{jadhav2012cyborg}. }
\label{fig:false-color2}
\end{figure}

\section{Future Directions}

In this work, several intrabody devices were reviewed from the perspective of their potential to be the sensor nodes of the future Internet of Bio-Nano-Things. These devices are, up to this moment, the smallest cell-stimulators, that can be applied to the monitoring of human health state or to any kind of disease treatment. Much work still need to be done within this field. Several challenges, which are detailed in the following section, need to be addressed before implementing a fully-fledged nanonetwork. At the same time, the current technology level is mature enough and have potential to evolve into the Internet of Bio-Nano Things vision. Here we also show some of the possible outcomes that will be enabled by this technology.

\subsection{Challenges}

The state-of-the-art presented in this paper shows that most of the devices are in the milimeter or micrometer scale and have limited functionality. Nevertheless, they have the potential to be further integrated into intrabody networks and connected with the Internet which will enable the development of biotechnological solutions based on the real-time monitoring and controlling of the human health through intrabody nanoscale devices. Other challenges hinder the implementation of an Internet of Bio-Nano Things dedicated to health applications and they are outlined in this section.

--- \textit {intrabody devices size.} The state-of-the-art is composed by devices with dimensions in the millimiter and micrometer scale. Depending on the application, for example for neural brain machine interfaces, the smaller the better to reduce the patient discomfort. Moreover, smaller intrabody devices will enable a more focused (cell-size) sensing and actuation, reducing any side effects to the neighbour cells.

--- \textit {Powering the devices.} intrabody devices need to have a power source capable of operate for long periods or that are rechargeable to reduce or even avoid their removal from the patient to replace batteries. Therefore, scientists have been investigating methods for energy harvesting and wireless power transfer to power the intrabody devices \cite{donohoe2017,rajavi2017rf,stehlin2012implantable, chow2010fully,charthad2015mm,maleki2011ultrasonically,cheng2015wireless}. Nevertheless, these solutions do not efficiently deliver power to intrabody devices, resulting in limited power supply. Moreover, these power sources have to be placed close to the device that they are powering to lower the collateral damage that could happen higher power signal are used. Currently, from the devices described in this paper, the longest distance from the implant to the power source is 5 cm. In order to operate, most of the implants require to have the energy capacitor close to the skin. The intrabody devices with non-rechargeable battery are expected to be taken from the body in a really short period. This challenges introduce inability of long-term wearing the implants and constantly monitoring the physiological parameters, as well as limit the functionalities of the implant.  

--- \textit{Data transmission interference.} The second issue concerns the way that data communication is performed over long distances with low power consumption. Inductive, optical and radio links have been used to establish a data link between the implanted device and the external device. Taking into account that devices which are reviewed here are operating with a similar frequency range (co-channel) - it can introduce additional interference that could corrupt the data collected by the intrabody devices. Currently, the most common operating frequency is 2.4 GHz, which is the Industrial, Scientific and Medical band - ISM band (standard for these type of applications) \cite{shon2017implantable}. Further research is needed to improve the communications links among intrabody devices and their with any external device, which will connect this sensor network to the Internet, and counter the co-channel interference.

--- \textit{Material composition.} The composition of the intrabody devices is important to ensure bio-compatibility. Currently, the intrabody devices have been made using piezoelectric materials, platinum, gold, titanium, copper, silicon, silver, aluminum. Apart from those, one material that has been investigated for this area is the graphene \cite{li2008}. This material enable the construction of very thin devices, up to a few nanometers, which have good electrical and thermal conductivity, in addition to be stronger than steel \cite{li2008}. A scientist from FGBU, Russia have performed experiments related to graphene compatibility and have shown that its asserts no influence on animal's health (has been tested on rats) \cite{savicheva2018Experiment}. It has been also shown the evaluation of the development of microorganism on a graphene plate (the experiments has been performed with \emph{Staphylococcus aureus}, \emph{Escherichia coli} and \emph{Streptococcus agalactiae}) \cite{savicheva2018Experiment}. The microorganisms have shown no growth, despite no antibiotic features have been found in graphene during this investigation.

--- \textit{Security.} Security is a important matter when developing human-machine interaction solutions. To secure the personal information an infrastructure for body area network need to be implemented. The security mechanisms must be much stronger than they currently are. In a study published in 2017, a research group reviewed several examples of how remote controlled intrabody devices are unreliable. In the examples, they included a group that was able to hack an insulin pump, an undesired electrical shock delivered to a cardiac pacemaker, and a successful attempt of hacking a number of different types of medical implants in 2016, under black box testing conditions \cite{zheng2017wannacry}. The lack of encrypted communications for the current intrabody devices is also a threat to the data security and integrity. Most of these systems do not address security issues as the countermeasures result in extra resources consumption, which in some cases an external security module need to be applied, resulting in bigger intrabody devices. Therefore, the further development of this field is dependent on the advancements of the Internet of Bio-Nano Things security research.

--- \textit{Timely cell detection}.  A cell being under mechanical or chemical influences can be deformed and change its characteristics leading them to incorrectly inform about their biological structure and functions. The size of biological cells is range from $1\,to\,100\,\mu\text{m}$. The cellular cytoskeleton define its resistance against deformation. Any influence on cell's cytoskeleton can corrupt the data that the cell stores or even worse - provoke a diseased state \cite{bao2003cell}. Considering this, one must be very careful when storing cellular information, as some data might be not in accordance with the cell real condition. Besides this, the cell can die before being detected, which will cause the change of the molecular network and affect consequently the overall received signal \cite{ingber1998search}. Moreover, all the interventions should not inappropriately induce the cellular apoptosis (i.e. necrosis) to prevent in the future grow of tumors \cite{chen1997geometric}. Apoptosis is a vital process of developments of organism including normal cell turnover, proper development and functioning of the immune system, etc \cite{elmore2007apoptosis}. Inappropriate apoptosis can become induce several diseases, including neurodegenerative diseases, ischemic damage, autoimmune disorders and many types of cancer.

\subsection{Perspective}

The medicine and bio-engineering are developing in a fast pace. The expectation is the next years a new technological leap on the biotechnological area would bring it closer to medicine, resulting in novel theranostics for humans. At this stage, intrabody devices would become pivotal for the implementation of real-time health-oriented sensor networks that would support the treatment of chronic diseases, as well as for the early detection of cancer and novel viruses. Here we list a few opportunities that arise from the further integration of intrabody devices using bio-cyber interfaces.

--- \textit{DNA control and gene therapy.} The behaviour of DNA molecules can be controlled by applying an electromagnetic field to gold nanocrystals that are covalently linked to these DNA molecules \cite{hamad2002remote}. Through this inductive link, the DNA molecules get agitated, increasing its temperature, and thereby changing its structure. The nanocrystals act as nano antennas (1.4 nm of length) and produce highly localized inductive heating when stimulated using an electromagnetic signal with the frequency of 1 GHz \cite{hamad2002remote}. This heating process separate the DNA into two strands in a matter of seconds in a fully reversible dehybridization process that leaves neighboring molecules untouched. In this scenario, bio-cyber interfaces can support the monitoring of specific parts of the human body and control the delivery of biochemical signals to control these specific cell's DNA behaviour. Moreover, a remote and autonomous gene therapy can be orchestrated using bio-cyber interfaces that monitor the health state of single or multiple cells. A simplified version of this approach using magnetic nanoparticles was proposed in 2016. Here we would like to emphasize that in the future, interconnected biocompatible devices will enable such solutions for local actuation, autonomous behaviour, and remote monitoring of the cell state.

--- \textit{Clinical use Body Area Nanonetwork to treat diseases.} The research interest in body area networks arose tied with the development of the microelectronics with embedded computing power and wireless communications systems. Using these networks, physiological data is collected from the person's body and exchanged among the sensor nodes through wireless links. Similarly, in the future, nanoscale sensing devices, such as the ones described in Section \ref{sec:intrabodybio}, will be interconnected in nanonetworks limited to each person's body area and will exchange information using the Internet infrastructure. In this scenario, each person could have several body area nanonetworks, where each one of them would be specialized in a particular area of the patient's body or even in a particular health condition. This medical treatment would be prescribed by physicians, which would identify and fit the body area nanonetwork for the patient needs, and designed by biomedical engineers, which would follow the physician's prescription to build the most effective body area nanonetwork.   

\section{Conclusions}

In the next years, the further advancements in medicine, nanoelectronics, molecular communications and biomedical engineering will pave the development of novel theranostics that will be personalized to the patient's health characteristics and based on its physiological data collected by intrabody nanoscale devices. The pervasiveness of this approach will be dependent on the Internet infrastructure to enable the remote and real-time management of the intrabody nanoscale devices required for the reliable operation of this system. Therefore, the evolution of these nanoscale devices into bio-cyber interfaces is a necessary step towards the fully-fledged realization of this technology, where several nanonetworks would be interconnected and seamlessly controlled through the Internet.

\bibliographystyle{ieeetr}
\bibliography{sample}

\end{document}